\documentclass[reprint,superscriptaddress,
showpacs,
 amsmath,amssymb,
 aps,
]{revtex4-1}

\usepackage{graphicx}
\usepackage{dcolumn}
\usepackage{bm}

\usepackage{subfigure}
\usepackage{graphicx}
\begin{document}


\title{A quantum algorithm for greatest common divisor problem}

\author{Wen Wang}
\affiliation{School of Physics, Peking University, Beijing 100871, China}

\author{Xu Jiang}%
\affiliation{School of Physics, Peking University, Beijing 100871, China}


\author{Liang-Zhu Mu}
\email{muliangzhu@pku.edu.cn}
\affiliation{School of Physics, Peking University, Beijing 100871, China}

\author{Heng Fan}
\email{hfan@iphy.ac.cn}
\affiliation{Institute of
Physics, Chinese Academy of Sciences, Beijing 100190, China}
\affiliation{School of Physical Sciences, University of Chinese Academy of Sciences, Beijing 100190, China}
\affiliation{Collaborative Innovation Center of Quantum Matter, Beijing 100190, China}



\date{\today}

\begin{abstract}
We present a quantum algorithm solving the greatest common divisor (GCD) problem. This quantum algorithm possesses similar computational complexity
with classical algorithms, such as the well-known Euclidean algorithm for GCD.
This algorithm is an application of the quantum algorithms for the hidden subgroup problems, the same as Shor factoring algorithm.
Explicit quantum circuits realized by quantum gates for this quantum algorithm are provided.
We also give a computer simulation of this quantum algorithm and present the expected outcomes for the corresponding quantum circuit.

\end{abstract}

\pacs{03.67.Ac, 03.67.Lx}
\maketitle


\emph{Introduction.}---
The quantum algorithms may perform more quickly in solving certain problems than the classical ones for the same problems
in classical computers. The most famous one is Shor algorithm\cite{shor}, which may attack 
the existing cryptographic system involving Rivest-Shamir-Adleman (RSA) protocol\cite{RSA}.
The implementation of Shor algorithm in a quantum computer is still a tough task, however, great progresses have been made
experimentally. A realization of factoring $15$ into $3$ and $5$ using Shor algorithm was completed 
using room temperature liquid-state nuclear magnetic resonance techniques \cite{Chuang}. 
Recently, this task was also completed by using an ion-trap quantum computer \cite{15}. 
Since the number of available qubits which can be controlled is limited, the Kitaev approach\cite{Ki} 
to compress the implementation of Fourier Transform has been used, 
and controlled multipliers in the original algorithm is replaced 
with maps considering possible emergent states in the process of the algorithm.
However, this process costs much larger computational complexity than classical algorithms, 
diminishing the advantage in implementing the Shor algorithm. 
Many others have also investigated circuits containing as few qubits as possible 
though involving more gates \cite{Beau,in}. 
An adiabatic quantum algorithm is proposed and implemented to solve the factoring problem, 
succeeding to use fewer qubits than standard Shor algorithm \cite{DU1,DU2}. 
Four photonic qubits are used to complete the factorization of $N = 15$, 
coherently implementing the quantum circuits of the modular exponential execution and semiclassical 
quantum Fourier transformation \cite{PAN}.

In Shor algorithm, the true modular multipliers are necessary for 
the quantum advantage, though difficult to implement presently. 
However, we know that the modular addition may be realized more easily by current
technology. An quantum algorithm constituted by modular addition as a key step will be
of interest. Particularly, more effective quantum algorithms are in demand considering
large-scale quantum computers may be available in the near future.   
In this Letter, we present this quantum algorithm to solve the GCD problem (greatest common divisor problem). 
Similar to the Shor algorithm, this algorithm can be considered as
belonging to quantum algorithms for hidden subgroup problem \cite{QI1,QI2}.

The computational complexity of our quantum algorithm is $(\log_2(n))^2$. 
It shares similar time complexity with the classical algorithms for working out GCD.  
The existing classical algorithms include the well-known Euclidean algorithm \cite{EU,QI2} and Stein algorithm \cite{ST}. 
The complexity of the latter is $O((\log_2(n))^2$) and the complexity of the Euclidean algorithm is $O((\log_2(n))^3)$. 
We remark that our algorithm needs some efficient classical algorithm such as Stein algorithm to deal with the final result of the measurement, 
or it would need to iterate the algorithm for extra $O(\log_2(n))$ times to get the GCD. In that case, the complexity of our quantum
algorithm would grow to $O((\log_2(n))^3)$. Shor factoring algorithm solves a problem that there does not exist a classical algorithm
which can work out effectively. But for the GCD problem, the existing classical algorithms work efficiently enough. 
So our quantum algorithm would not give such advantages as Shor factoring algorithm, but comparable with the classical ones. 
However, this one is simpler than the Shor algorithm, even though their complexities are the same. 
So it can act as a test algorithm for quantum computers since it is a typical application of the quantum algorithms for the hidden subgroup problems.

\emph{Mathematical foundation.}---
In the GCD problem, our purpose is to obtain the greatest common divisor of two given natural numbers, say $x$ and $r$. We can do it by looking for the least positive integers $N$ and $k$ to establish the relationship $Nx=rk=P$ in which $P$ is an integer. At this point we know that $P$ is the least common multiple of $x$ and $r$ and that $\frac{r}{N}$ is the greatest common divisor of $x$ and $r$. Our quantum algorithm is designed to obtain the number $N$ so that we obtain the $\frac{r}{N}$. To be specific, we will obtain an estimation of $\frac{s}{N}$ in which $s$ is a random number satisfying $0\leq s \leq (N-1)$ with enough sufficient accuracy by the end of the quantum procedure. Combining with some other classical calculation and repetition of the quantum procedure, we can obtain the final result $\frac{r}{N}$. We have two different methods to do this conversion from $\frac{s}{N}$ to $\frac{r}{N}$ and they are both discussed in the section ``Further processing''. 
And after we obtain the $\frac{r}{N}$, the problem gets solved.

\emph{Quantum procedure.}---
We use the modular addition in our algorithm as the controlled unitary operator.
\begin{equation}
\begin{aligned}
U : |y\rangle \rightarrow |(x+y)\text{mod} r\rangle
\end{aligned}
\end{equation}
Here if we operate the state in $r$ dimensional space, the operation ``mod$r$'' could be satisfied automatically. If not, we use some quantum gates to do so. Then we claim that for any integer $s\in{0,1,...,N-1}$ , the state $|u_s\rangle$ is the eigenstate of U.
\begin{eqnarray}
\label{eq:us}
U |u_s\rangle &=&
U ( |y\text{mod}r\rangle +e^{-2\pi i \frac{s}{N} \cdot 1} |(x+y)\text{mod}r\rangle \nonumber \\
&&+ e^{-2\pi i \frac{s}{N} \cdot 2} |(2x+y)\text{mod}r\rangle + ... \nonumber \\
&&+ e^{-2\pi i \frac{s}{N} \cdot (N-1)} |((N-1)x+y)\text{mod}r\rangle)\frac{1}{\sqrt{N}}\nonumber \\
&=&e^{2 \pi i \frac{s}{N}} |u_s\rangle
\end{eqnarray}
Here the parameters $x,r,N$ have already been defined previously. 
So we have found the eigenstate of $U$, and we can use the techniques of quantum phase estimation \cite{QI1,QI2} to obtain $N$. The procedure is as follows,
\begin{itemize}
\item $|0\rangle^{\otimes t} |0\rangle $
\item $\rightarrow \frac{1}{\sqrt{2^t}}\sum_{j=0}^{2^t-1} |j\rangle |0\rangle $  \hfill apply $H^{t}\otimes I$
\item $\rightarrow \frac{1}{\sqrt{2^t}}\sum_{j=0}^{2^t-1} |j\rangle | (jx) mod r \rangle $  \\
$= \frac{1}{\sqrt{N2^t}}\sum_{j=0}^{2^t-1} \sum_{s=0}^{N-1}  |j\rangle | u_s \rangle e^{2 \pi  i s j/N}$ \hfill apply $c-U$
\item $\rightarrow \frac{1}{\sqrt{N}} \sum_{s=0}^{s=N-1} |\widetilde{s/N} \rangle | u_s \rangle $  \hfill apply  $F^+ \otimes I$
\end{itemize}
Here  $y$ in the $| u_s \rangle$ has been assigned as 0. And here we have used  $|jx+y\rangle=\frac{a}{\sqrt{N}}\sum_s e^{2\pi i s j/N} |u_s\rangle$. This result can be easily obtained from Eq.(\ref{eq:us}). $F$ is the quantum Fourier transform and $F^+$ is the inverse quantum Fourier transform. The definition of $|\widetilde{\psi} \rangle$ is as follows, if $\varphi_u = 0.\varphi_1 \varphi_2
\varphi_3 ...$, $\widetilde{\varphi_u}=\varphi_1 \varphi_2 \varphi_3 ...$, where $\varphi_i \in F_2$ is the number on the $i$'th position of the binary number $\varphi_u$.

\begin{figure}
\centering
\subfigure[The circuit for the first two operations in the quantum procedure.]{
\label{fig:subfig:a} 
 \includegraphics[width=0.48\textwidth]{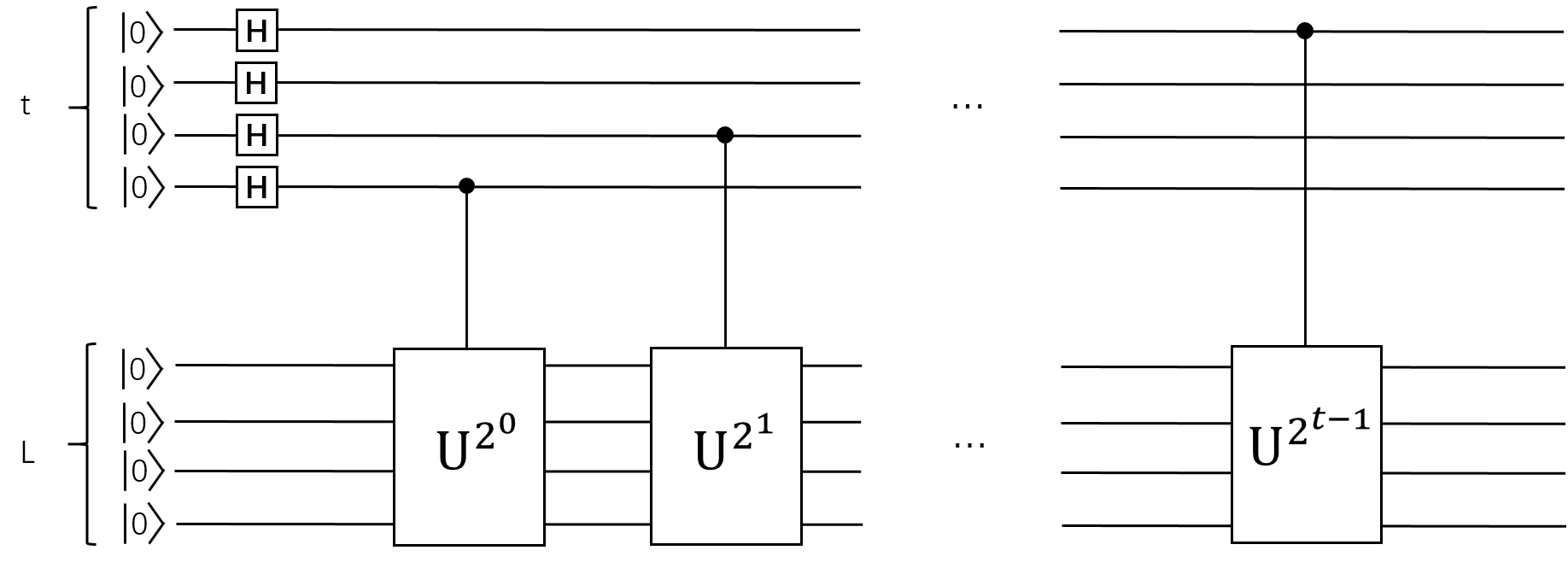}}
\hspace{1in}
\subfigure[The implementation of the $U^{2^i}$. Here we use the techniques of reversible computation\cite{QI2}.]{
\label{fig:subfig:b} 
\includegraphics[width=0.4\textwidth]{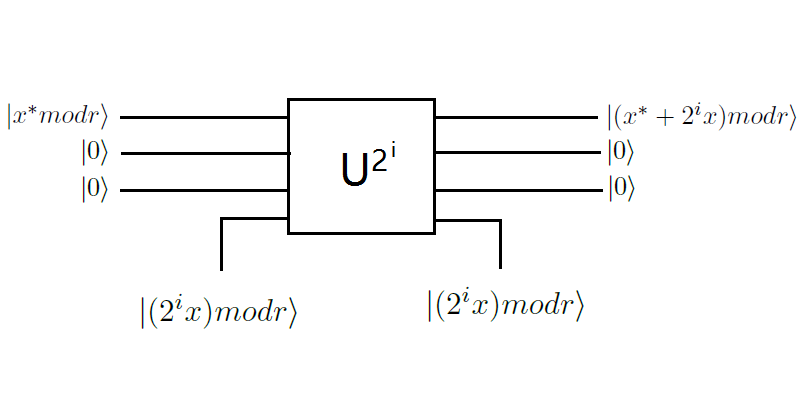}}
\caption{A practical method of implementation of the first two operations in the quantum procedure.}
 \label{fig:1} 
\end{figure}

According to the theory of the quantum phase estimation, if we choose
\begin{equation}
\begin{aligned}
t= n + \left\lceil \log _2(2+\frac{1}{2 \epsilon})\right\rceil
\end{aligned}
\end{equation}
we can obtain an estimation of $s/N$ accurate to $n$ bits with a success rate of at least $1-\epsilon$. We may use the continued fractions algorithm\cite{CFA} just like Shor algorithm to obtain the precise $s/N$ from the estimation. In that case $n$ needs to be at least $2L+1$. But because our problem is different, we are able to make $n$ a smaller one, which makes implementation easier. Let $b$ be the estimation of $s/N$ accurate to $n$ bits and $b^{'}=\frac{s}{N}-b$. We calculate $p = \left\lceil b\times r-\frac{1}{2} \right\rceil$. We know $r/N$ is an integer. So we can say $p/r=s/N$ only if $b^{'}$ satisfies
\begin{equation}
\begin{aligned}
r \times | b^{'} | \leq \frac{1}{2}
\end{aligned}
\end{equation}
Considering $r<2^{L}$ when $L=\left\lceil \log _2r \right\rceil$, $n$ only needs to satisfy $n\geq (L+1)$. So compared with the least $t=2L+1+\log _2(2+\frac{1}{2\epsilon})$ in Shor algorithm, this algorithm only needs $t=L+1+\log _2(2+\frac{1}{2\epsilon})$ and we could get the precise $p/r=s/N$.

\emph{Further processing.}---
However, we only have obtained $p/r=s/N$ the numerator and denominator of which may share a common factor. We still need to use other classical algorithms to simplify it to 
find a coprime ratio, for example, by using the Stein algorithm. Since we could directly use Stein algorithm to solve the whole problem, this quantum algorithm performs not as well as this classical one. We also know $s$ and $N$ may share a common factor. Takeing this situation into account, we repeat the quantum procedure for a constant number of times and pick the greatest $N$. Let $m$ be the total number of times to repeat, the probability of success of obtaining the true $N$ is more than $1-(\frac{3}{4})^m$\cite{QI2}. We define gcd$(x,r)$ as the greatest common divisor of $x$ and $r$. And then we could obtain the result gcd$(x,r)=r/N$.

If we do not look for a classical algorithm for help to find the coprime ratio, we can also complete the whole algorithm by repeating the quantum procedure. We choose the smaller one of $r$ and $x$ and define it as $x^{'}$, and we choose the smaller one of $s/N\times r$ and $x^{'}-s/N \times r$ and define it as $r^{'}$. We use the new $x^{'}$ and $r^{'}$ to replace the given $x$ and $r$ in the quantum procedure and repeat the procedure. Every time we repeat the quantum procedure and obtain a result $s_{i}/N_{i} \times r_{i}$. We need to test if it is the common divisor of $r$ and $x$. If it is, we claim that the result is the gcd$(x,r)$ and the nonzero result will not change again if we continue the repetition. If not, further repetitions are needed to continue. Totally, the number of repetitions is $O(\log _2(n))$.

Here we need to notice that the method discussed above works successfully only if gcd$(s_{i}/N_{i}\times r_{i}$,$x_{i+1})=\text{gcd}(x_{i},r_{i})$. That means, for example, if $r_{i}$ is lager than $x_{i}$ so that $x_{i+1}=x_{i}$, $s_{i}$ needs to share no common factor with $x_{i+1}/r_{i} \times N_{i}=k_{i}$. Here we have already considered that gcd$(a,b)=$gcd$(a-b,b)$. But because $s_{i}$ is a random number smaller than $N_{i}$, it maybe bring an extra factor to the result $s_{i}/N_{i}\times r_{i}$ if it shares a common factor with $k_{i}$. And gcd$(s_{i}/N_{i}\times r_{i},x_{i+1})$ would be larger than gcd$(x_{i},r_{i})$. Considering this situation our strategy is that when we find the nonzero result do not change again but it is not the factor of either of $x$ and $r$, we pick the one of $x$ and $r$ which is not divisible by the result as the $x_{i+1}$ for the next repetition. This would continue until we obtain a result that could be a factor of both $x$ and $r$.
This will not add much to the number of repetitions, since every time we still get a result that is smaller than half of the previous result. So the number of repetitions is still $O(\log_2(n))$.

\emph{Complexity.}---
In the quantum procedure we consider the time complexity. The Hadamard gates cost $O(1)$. For the modular addition shown in FIG. \ref{fig:1}, the classical calculation of $2^{j}x$ for all $j$ cost $O(\log_2(n))\times t=O((\log_2(n))^2)$, since we could calculate an addition $2^{j}x=2^{j-1}x+2^{j-1}x$ for each $j$ and each addition cost $O(\log_2(n))$, and the quantum circuit totally costs $O((\log_2(n))^2)$ gates containing $t$ $U^{2^i}$ operations each of which costs $O(\log_2(n))$ for modular addition. So this step costs $O((\log_2(n))^2+O((\log_2(n))^2=O((\log_2(n))^2$. The inverse Fourier transform costs $O((\log_2(n))^2$. So the quantum part costs $O((\log_2(n))^2$. If we use a classical approach like Stein algorithm to deal with the result of measurement. Finally we need $O((\log_2(n))^2$. 
If we choose to repeat the quantum procedure, the total number of repetitions that are needed is $O((\log_2(n))$ and the total complexity of the algorithm grows to $O((\log_2(n))^3$.

\begin{figure}[h!]
    \centering
    \includegraphics[width=8.6cm]{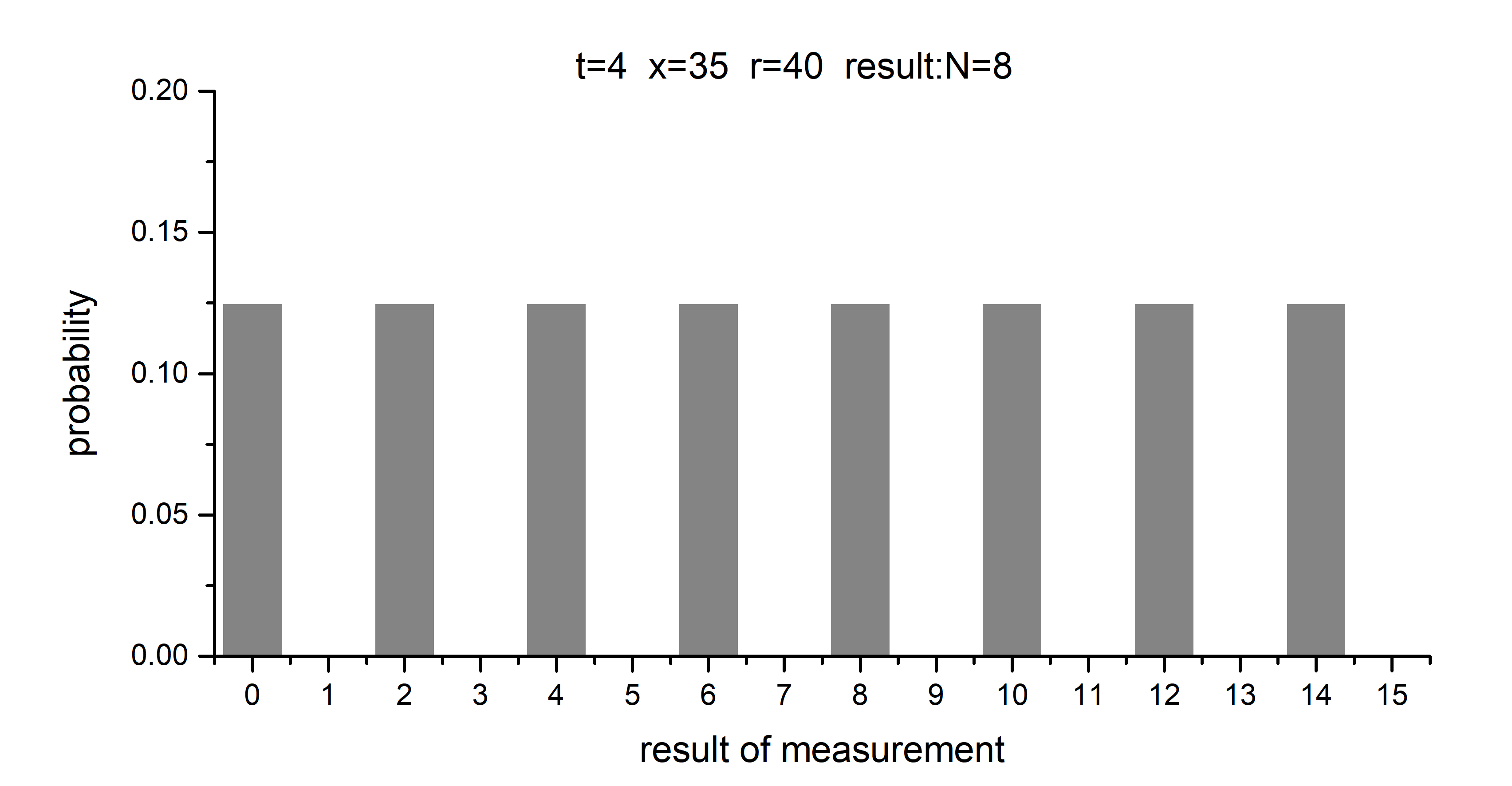}
    \caption{\label{fig:8}
a computer simulation of the quantum procedure in our algorithm, with $t=4, x=35, r=40$ so that $N=8$. The possible results of measurement share the same probability of $0.125$ and are distributed evenly. The results of measurement are $b\times 2^{t}$. For example, if our detection gives a result of $2$. And here we have $t=4$ so $b\times 2^{4}=2$ and $b=0.125$. After calculation we obtain $s/N=0.125$. If we use the stein algorithm we could obtain $N=8$. If we choose to repeat the quantum procedure we could determine that for the next repetition $x^{'}=s/N\times r=5$.
		}	
\end{figure}

\begin{figure}[h!]
    \centering
    \includegraphics[width=8.6cm]{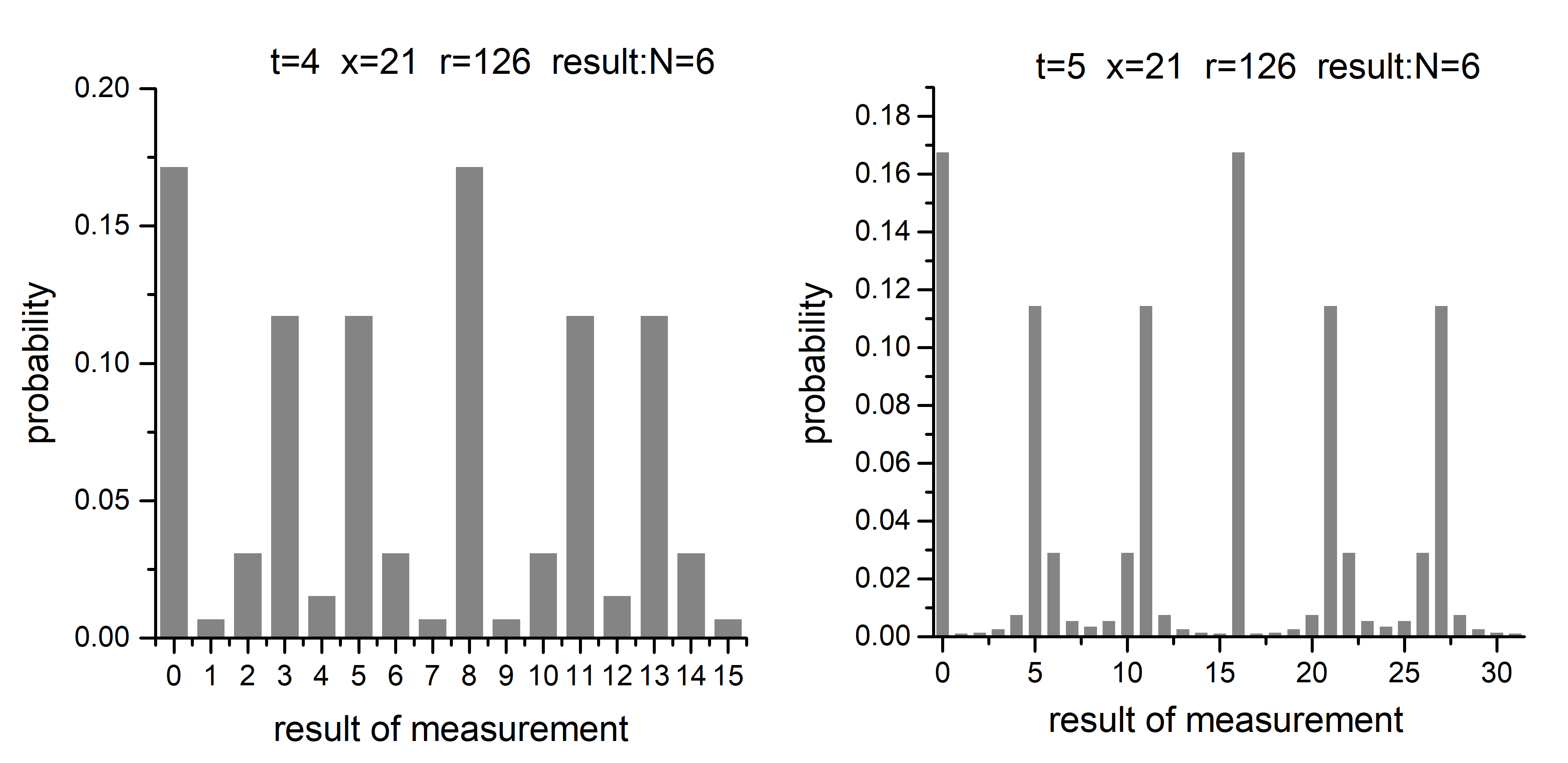}
		\includegraphics[width=8.6cm]{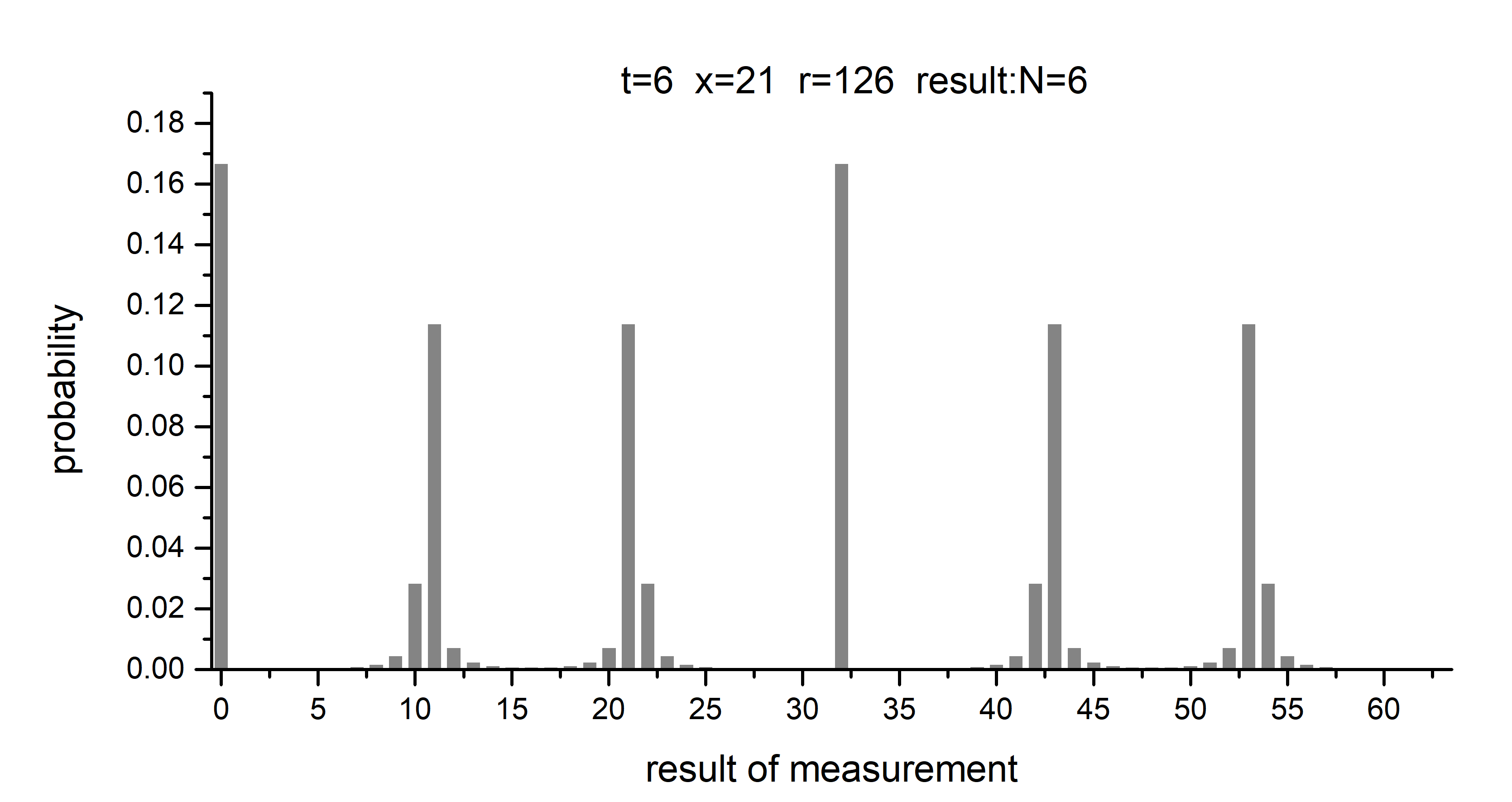}
    \includegraphics[width=8.6cm]{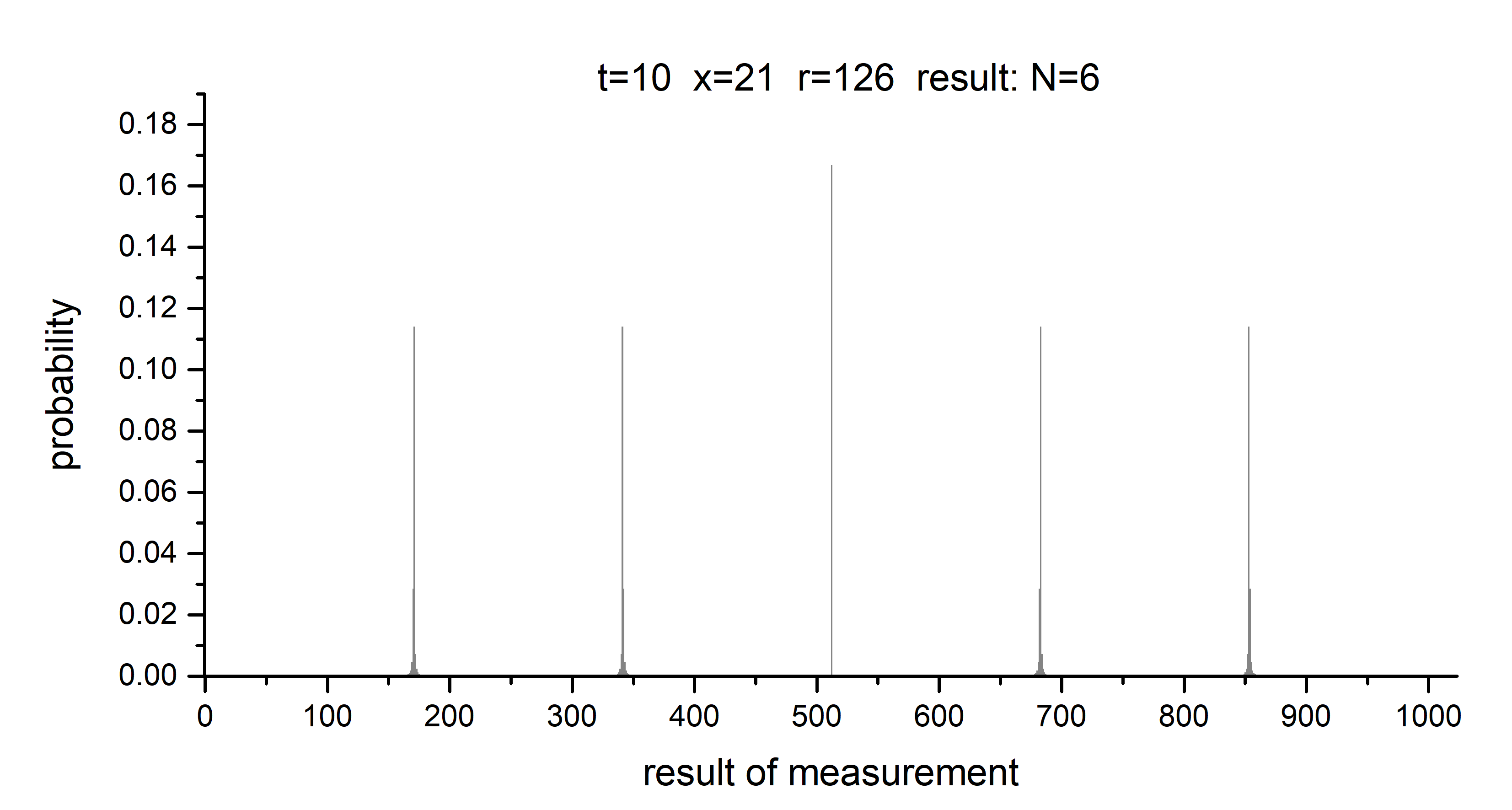}
    \caption{\label{fig:6}
a computer simulation of our algorithm, with $t=4/5/6/10, x=12, r=126$ so that $N=6$. The peaks are distributed evenly but have slightly different probability. The results of measurement are also $b\times 2^{t}$.
		}	
\end{figure}

\emph{Simulation of the algorithm.}---
In this part we give two samples of the computer simulation of our algorithm. In the first one we choose $x=35, r=40$ so $N$ of this problem should be $8$, shown in FIG. \ref{fig:8}. We can see eight uniformly distributed peaks of the probability distribution, each of which has a probability of $\frac{1}{8}$. In the second one we choose $x=21, r=126$ so $N$ of this problem should be $6$, shown in FIG. \ref{fig:6}. We can see six uniformly distributed peaks of the probability distribution. As the total number of steps increase, they become more and more concentrated. These two samples show that their behaviors are just the same as Shor algorithm \cite{QI2,in}. So it presents that our algorithm gives a same structure of results as Shor algorithm.

\emph{Discussion.}---
The quantum algorithm presented in this Letter is slightly simpler for implementation than the Shor algorithm. In addition, the result also gives a periodic structure just like the complex Shor algorithm. The heart of this algorithm is the application of modular addition, which is known as a basic element of modular exponentiation. 
So its implementation can act as a prelude before we have sufficient ability to implement the full controlled modular exponentiation. 
We do not need to design other operations to replace the part of modular addition. And we can analyze the experimental results of this kind of quantum algorithms solving the hidden subgroup problem in an easier way.

\begin{figure}[h!]
    \centering
    \includegraphics[width=8.6cm]{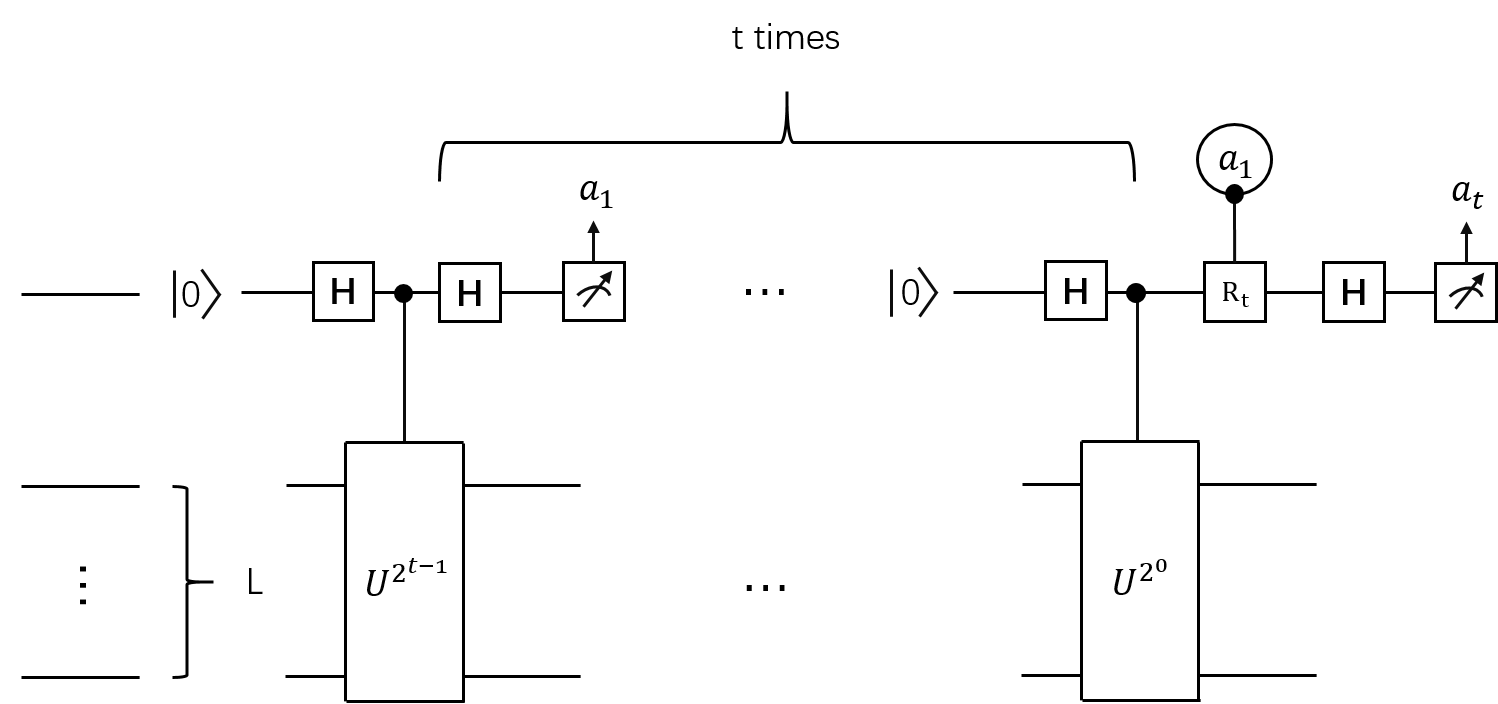}
    \caption{\label{fig:ki}
		The Kitaev's approach to reduce the number of qubits in the quantum circuit.
		}	
\end{figure}

\begin{figure}[h!]
    \centering
    \includegraphics[width=8.6cm]{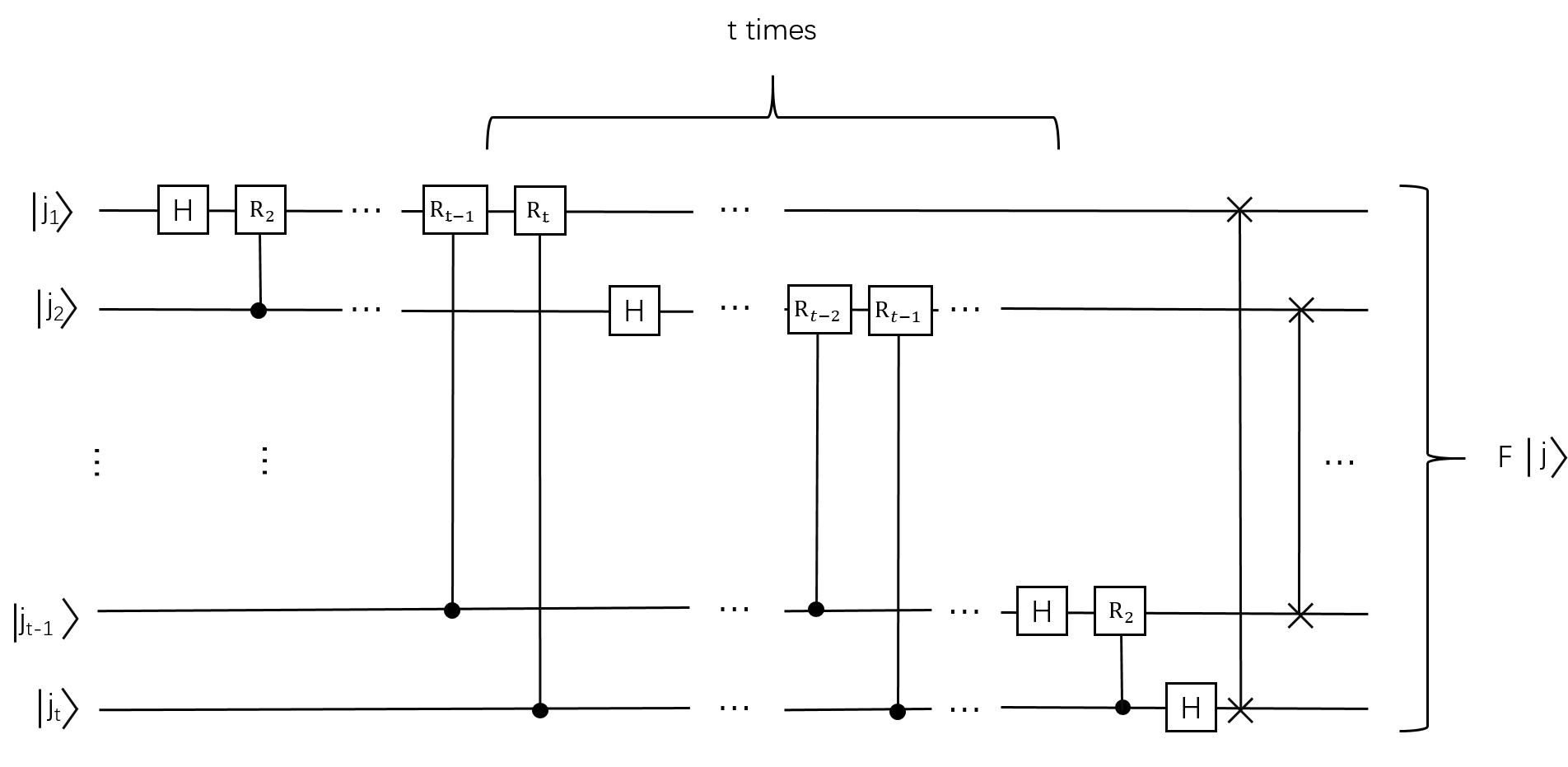}
    \caption{\label{fig:FU}
		A efficient circuit for quantum Fourier transform. The inverse quantum Fourier transform is just the inverse one of this circuit.
		}	
\end{figure}

The implementation of modular multiplier and modular addition considering different circumstances can be seen in many previous works\cite{Beau,in,VED,FAST}. 
In the scheme \cite{VED}, the total number of gates for modular multiplier is $5L+3$, while that for modular addition is $4L+2$, which means we need less qubits than Shor algorithm to get a same result structure as the Shor algorithm. In addition, we have demonstrated that in this algorithm, $t=L+1+\log_2(2+\frac{1}{2\epsilon})$, 
while in Shor algorithm, $t=2L+1+\log_2(2+\frac{1}{2\epsilon})$. It means we need a smaller number of qubits and less gates. 
It also means that we only need about a half of steps to obtain a result with high enough accuracy. The matrix form of phase gates is as follow,
\begin{equation}
\begin{aligned}
R_t=\begin{bmatrix} 1 & 0 \\ 0 & e^{2\pi i/2^t} \end{bmatrix}
\end{aligned}
\end{equation}
Considering Kitaev's approach \cite{Ki} shown in FIG. \ref{fig:ki}, the smallest angle that appears in the phase gates is $\pi/2^{t-1}$. 
It means that a larger $t$ leads to a requirement of higher accuracy of the implementation of the phase gates, 
the same as in the quantum inverse Fourier transform shown in FIG. \ref{fig:FU}. 
Thus, this quantum algorithm is a test algorithm for a quantum computer before we have the ability to complete the full implementation of Shor algorithm.

In conclusion, we give a quantum algorithm solving the greatest common divisor problem. Its computational complexity is similar as classical algorithms. 
We have given the explicit quantum circuit and the processing schemes for the results of the measurements. We also give the computer simulations of the algorithms and analyze the probability distribution of the expected outcomes of the circuit. Finally we discuss the advantages of our algorithm and show that it is a good test algorithm 
for quantum computers solving the kind of the hidden subgroup problems.

\emph{Acknowledgements:} This work was supported by the National Key R \& D Plan of China (No. 2016YFA0302104,
No. 2016YFA0300600), the National Natural Science Foundation of China (No. 91536108),
and the Chinese Academy of Sciences (No. XDB01010000, and No. XDB21030300).

\end{document}